\documentclass[12pt,letterpaper]{article}
\usepackage{epsfig}                 
\usepackage[authoryear]{natbib}
\usepackage{graphicx}
\usepackage{amsmath}
\usepackage{psfrag}
\usepackage{mathabx}

\large
\usepackage[usenames,dvipsnames]{color}
\usepackage{fullpage}
\usepackage{multirow}

\usepackage{color}
\usepackage[normalem]{ulem}  
\title{Scrambling eggs: Meiotic drive and the evolution of female recombination rates.}
\usepackage{fancyhdr}
\author{Yaniv Brandvain \\ email: ybrandvain@gmail.com  \and Graham Coop \\ email: gmcoop@ucdavis.edu }
\date{}
\begin{document}
\maketitle
\begin{center}
Center for Population Biology \& Section in Evolution and Ecology \\ University of California - Davis \\ Davis, CA, 95616
\end{center}

\newpage

{\bf Abstract:}
Theories to explain the prevalence of sex and recombination have long been a central theme of evolutionary biology.
Yet despite decades of attention dedicated to the evolution of sex and recombination, the widespread pattern of sex-differences in the recombination rate is not well understood and has received relatively little theoretical attention.
Here, we argue that female meiotic drivers - alleles that increase in frequency by exploiting the asymmetric cell division of oogenesis - present a potent selective pressure favoring the modification of the female recombination rate. 
Because recombination plays a central role in shaping patterns of variation within and among dyads, modifiers of the female recombination rate can function as potent suppressors or enhancers of female meiotic drive.
We show that when female recombination modifiers are unlinked to female drivers, recombination modifiers that suppress harmful female drive can spread. 
By contrast, a recombination modifier tightly linked to a driver can increase in frequency by enhancing female drive. 
Our results predict that rapidly evolving female recombination rates, particularly around centromeres, should be a common outcome of meiotic drive.  
We discuss how selection to modify the efficacy of meiotic drive may contribute to commonly observed patterns of sex-differences in recombination.

\newpage
\section{Background}
Recombination plays a critical role in both the production of viable gametes and population genetics processes. 
Structurally, chiasmata - the physical manifestations of crossovers - generate the tension between homologs often needed to ensure proper segregation during meiosis I.
The structural role of chiasmata is likely the mechanistic underpinning of the requirement in most species of at least one recombination event per chromosome.
In addition to the structural role of chiasmata, recombination also plays an important role in population genetic processes by generating novel haplotypes within populations. 
The production of novel haplotypes entails both the creation and separation of beneficial alleles (or allelic combinations). 
The balance between these opposing outcomes shape the adaptive value of recombination, a topic of much research \citep[e.g. ][]{Eshel1970,Feldman1996,Otto2002,Keightley2006,Barton2009}.

Recombination rates vary between species, individuals, and sexes \citep{Bell1982,Trivers1988,Burt1991,Lenormand2003,Lenormand2005,Lorch2005,Coop2007}.
We address the evolution of sex-difference in the recombination rate (i.e. heterochiasmy).
We first characterize general patterns of sex-differences in recombination rates.
After reviewing the major theories invoked to explain the evolution of heterochiasmy, we introduce the major argument of this manuscript - that the common and consistent sex-difference in the operation of meiotic drive can favor the evolution of heterochiasmy.
We conclude by considering the implications of our model in light of our current understanding of the genetic basis of recombination modification.

\subsection{Observations}

Four broad patterns describe sex differences in recombination rates. We describe these patterns and their consistency below.

\emph{The achiasmatic sex is heterogametic (The Haldane-Huxley Rule)}: When recombination is absent in one sex (i.e. it is achiasmatic), that sex is nearly always the heterogametic sex \citep[i.e. the sex bearing heteromorphic sex chromosomes -- ][]{Haldane1922,Huxley1928,Burt1991}. This observation represents more than 25 evolutionary independent origins of sex-specific achiasmy \citep{Burt1991}, and is observed under both male (e.g. \emph{Drosophila}) and female (e.g. \emph{Lepidoptera}) heterogamy, with very few known exceptions \citep{Davies2005}.

\begin{figure}
\begin{center}
\includegraphics[height=8cm]{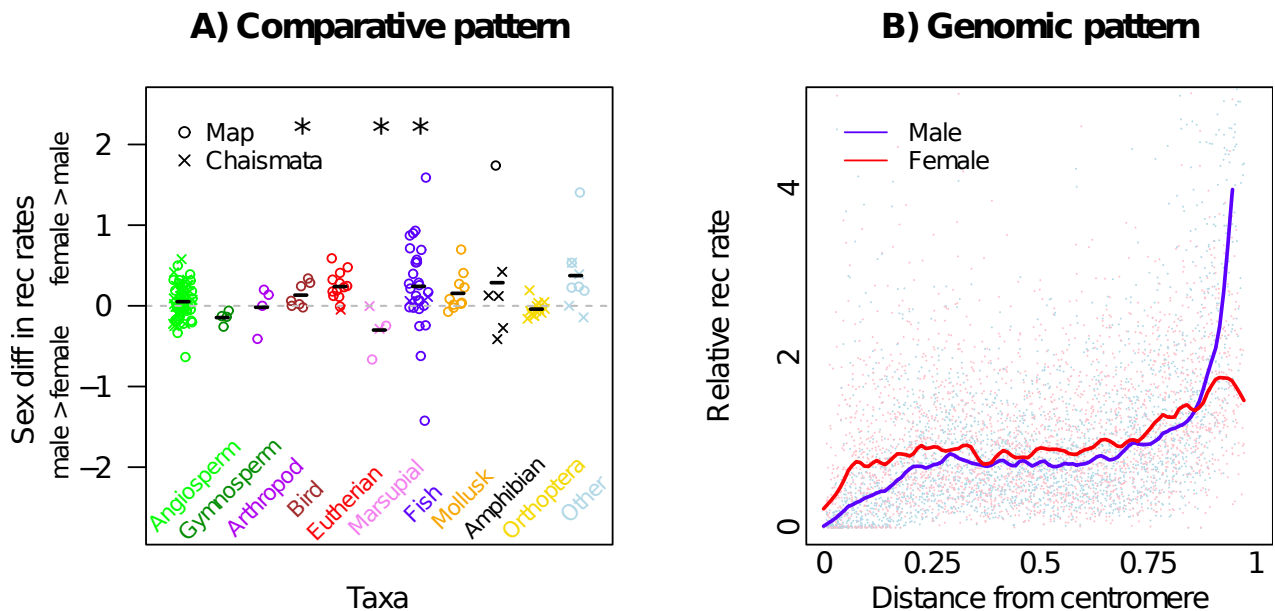} 
\end{center}
\caption{{\bf Genome wide and regional sex-differences in recombination rates.}
{\bf A)} The difference between female and male recombination rates ( $\circ$ from linkage maps, excluding known sex-chromosomes,  $\times$ from chiasmata counts) divided through by the sex-averaged rates.
Points above the dashed line indicate higher rates of recombination in females than in males.
$^*$ indicates $p < 0.05$ using a two-tailed sign test, without correcting for multiple tests or phylogeny, and ignoring ties.
{\bf B)}. Sex-standardized recombination rates across the human genome.
The sex-standardized rate equals the local recombination rate in a given sex (\textcolor{blue}{male} and \textcolor{red}{feamle}), divided by the average recombination rate in that sex.
The x-axis indicates the position of the focal genomic region (.2\% of a chromosome arm), divided by the length of the chromosome arm.
Data are presented from all metacentric human autosomes.
Lines represent a lowess smoothing of these points. }
\label{MaleVFemaleMaps}
\end{figure}

\emph{Females recombine more than males}: 
When both sexes recombine, the female recombination rate often exceeds the male rate \citep{Bell1982,Trivers1988,Burt1991,Lenormand2003,Lenormand2005,Lorch2005}. 
We display this pattern in Figure \ref{MaleVFemaleMaps}A by plotting the sex difference in recombination rates (measured by autosomal map lengths or chiasmata counts) in a number of taxa. We note that the grouping of our taxa is somewhat arbitrary; nonetheless, clear trends across groups in the degree of sex difference in recombination rate are apparent. 

The pattern of higher female recombination rates is quite broad, occurring in animal species with XY, ZW, and environmental sex-determination. 
However, there are many exceptions (e.g. marsupials, some grasshoppers, and newts) suggesting that the process of recombination is not mechanistically constrained towards higher rates in females and that the ratio of male to female rates can evolve quite rapidly (see \citet{Lenormand2005}). 
In plants, there is no clear trend towards higher recombination rates in female meiosis (Figure \ref{MaleVFemaleMaps}A); however, when outbreeding angiosperms are considered separately, recombination rates are on average slightly higher in female meiosis than in male meiosis \citep{Lenormand2005}.

\emph{Females recombine at relatively higher rates near centromeres}: 
After controlling for the genome-wide sex difference in recombination rate, females recombine more often near centromeres than do males, while males recombine relatively more near telomeres. 
This pattern has been noted in fish \citep{Sakamoto2000,Singer2002,Reid2007}, humans \citep{Broman1998,Kong2002,Clark2010}, dogs \citep{Wong2010}, and mice \citep{Shifman2006, Paigen2008}, although there are some exceptions \citep[e.g. opossums][]{Samollow2004,Samollow2007}. 
Utilizing data from a recent fine-scale analysis of sex-specific recombination rate in humans \citep{Kong2010}, we display an example of this pattern in Figure 1B. 

This pattern is not an obvious consequence of different requirements for segregation in males and females, as it is thought that recombination near centromeres does not contribute to cohesion of homologous chromosomes. 
Furthermore, recombination events too close to the centromere can generate problems during meiosis, such as an increase in the rate of precocious separation of sister chromatids, potentially leading to aneuploid gametes \citep[i.e. gametes with an abnormal number of chromosomes -- ][]{Rockmill2006}. 
The higher rates of recombination close to centromere in aneuploid transmissions \citep{May1990,Lamb2005} suggests that recombination close to the centromere may actually incur fitness costs due to errors during meiosis.

\emph{Genetic control of the recombination rate is often sex specific:}
The process of gametogenesis is fundamentally different in males and females. 
Consequently, the genetic control of many aspects of meiosis differs between the sexes \citep{Morelli2005}. 
Perhaps due to sex differences in meiosis, alleles that influence the recombination rate in one sex often have no influence (e.g. the 17q21.31 inversion region in humans), or the opposite effect (e.g. RNF212 in humans) on the recombination rate in the other sex \citep{Kong2004,Kong2008}. 
In fact, \citet{FledelAlon2011} show that there is no detectable heritable intersexual correlation in the recombination rate despite additive genetic variance in both sexes.

Although mechanisms governing sex differences in the recombination rate are not well-characterized, one candidate is the length of the synaptonemal complex (SC), a structure composed of cohesins and other proteins involved in meiosis, which stabilizes connections between homologs and facilitates crossing over. SC length is positively associated with the recombination rate, and is longer in females than in males \citep{Lynn2002,Tease2004,Dumont2011}.
Regardless of mechanism, the consequence of this pattern is clear - there is ample opportunity for sex-specific recombination rates in one sex to evolve independently of those of the other sex.

\subsection{Current theories}

Despite the large body of work on the evolution of sex and recombination \citep[for recent reviews see ][]{Otto2002,Barton2009}, the evolutionary forces responsible for the observed patterns of heterochiasmy have received curiously little theoretical attention \citep[but see ][]{Trivers1988,Lenormand2003,Lorch2005}. 
Below, we summarize the current of the evolution of heterochiasmy and review their plausibility. 

We note that these theories are not mutually exclusive, and it is unlikely that any one theory will explain all the observations presented above. 
Further, our model (like many of the other theories of the evolution of heterochiasmy) does not argue that selection directly favors the evolution of male and female rates in opposite directions. 
Rather, in our model, selection favors a modification in the female rate, with little or no direct selection on the male rate. 
In both our model and others, heterochiasmy results from the sex-specificity of recombination modifiers and/or additional selective constraints (such as stabilizing selection on the sex-averaged rate) which are not made explicit.
For example, if too much recombination is generally deleterious due to the costs of ectopic recombination, but circumstances favor an increase in female recombination rates, modifiers that increase female rates but not male rates may be favored, leading to the evolution of heterochiasmy.


\emph{Sex chromosome pleiotropy}: Based on their observations that the achiasmatic sex is heterogametic, \citet{Haldane1922} and \citet{Huxley1928} proposed that heterochiasmy may evolve as a pleiotropic consequence of selection to suppress recombination between heteromorphic sex chromosomes \cite[see also][]{Nei1969}. Although suppression of recombination between sex chromosomes could well explain the Haldane-Huxley rule, it cannot fully explain the quantitative variation in autosomal recombination rates between the sexes, as they span modes of sex determination, nor can this model explain variation across chromosomal regions. 

\emph{Maternal aging}: Physical connections between homologous chromosomes are necessary for proper segregation during meiosis. In many species, these physical connections are formed by chiasmata and therefore, one recombination event per chromosome is generally required for proper meiosis. 
In species where female meiosis is arrested, physical connections between chromosomes may degrade with time \citep[e.g. ][]{Lamb2005a}, and thus additional chiasmata may function to stabilize chromosomes across the metaphase plate.

According to the maternal aging theory, elevated female recombination rates provide more physical connections between chromosomes, ensuring that oocytes will segregate correctly after years of insults accumulated during meiotic pachytene arrest. 
This theory is supported by the finding that women with higher recombination rates have higher fertility  \citep{Kong2004} and that the viable gametes of older mothers have a higher number of crossovers, consistent with the idea that selection against those gametes with too few crossovers increases throughout a mother's reproductive life \citep{Kong2004, Coop2008}. 
However, this theory cannot easily explain heterochiasmy in organisms in which females create eggs throughout their lives.
Additionally, this theory cannot easily explain the spatial pattern of recombination in females, since elevated female rates are often accomplished by chiasmata placed in locations thought not to stabilize chromosomes.

\emph{Sexual selection}: \citet{Trivers1988} argued that selection to preserve high fitness genotypes favors recombination suppression in the sex with greater variance in fitness. 
Both current theory \citep{Lenormand2003} and data \citep{Burt1991,Mank2009} suggest that sexual selection cannot explain the evolution of heterochiasmy. 
Using multi-locus population genetic theory, \citet{Lenormand2003} showed that sex-differences in selection on diploid genotypes cannot generally favor the evolution of heterochiasmy. \citet{Burt1991} and  \citet{Mank2009} showed that the degree of heterochiasmy decreases with sexual-size dimorphism (a proxy for the strength of sexual selection) - an observation counter to predictions of the sexual selection hypothesis. 

\emph{Haploid and psuedo-haploid selection}: \citet{Lenormand2003} showed that simple sex differences in selection during the haploid life stage can favor the evolution of heterochiasmy. 
Like the sexual selection theory, the haploid selection model argues that the sex which produces the gamete (or gametophyte) with higher variance in fitness will recombine less. 
As sperm and pollen can experience intense competition for fertilization opportunities, \citet{Lenormand2003} argued that males should in general have lower recombination rates. 
\citet{Lenormand2003} also proposed that selection on haploid components of diploid genotypes (e.g. selection on epistasis in \emph{cis}, or imprinted loci) can also favor the evolution of heterochiasmy; we call this the pseudo-haploid selection model. 
Although theoretically plausible, it is unclear whether the small numbers of imprinted genes \citep{Morison2005} or genes expressed in sperm \citep[e.g. ][]{Joseph2004} are sufficient for the (pseudo-)haploid selection theory to explain heterochiasmy in animals. 
Furthermore, a comparative study found no association between heterochiasmy rates and the inferred strength of sperm competition in eutherian mammals \citep{Mank2009}. However, the absence of a haploid stage in the female gametes of animals and occasional haploid expression in sperm makes the haploid selection theory viable.

The situation is more complex in plants, as due to the alternation of generations, there is haploid expression in both the products of male and female meiosis. 
Nonetheless, \citet{Lenormand2005} argue that as in the majority of outcrossing plant species, selection is likely strong on male haploid products (due to pollen grain competition), the haploid-selection model could explain the female biased recombination rates in outcrossers.

\emph{Meiotic drive}: Below, we articulate the meiotic drive hypothesis. According to this model, sex differences in the operation of gametic drive presents a sex-specific selective pressure on linked and unlinked modifiers of the recombination rate.
Other authors have proposed that sex differences in meiotic drive may offer an opportunity for the evolution of sex-specific recombination rates \citep{Lenormand2003,Haig2010}. We discuss our work in relation to these models below.

\subsection{Meiotic drive and the evolution of heterochiasmy}
Meiosis provides an opportunity for alternative alleles to compete for representation in the functional gametes of heterozygotes. Alleles that distort meiosis and gametogenesis in their favor (i.e. gametic drivers) often do so at the expense of individual viability or fertility. Therefore, although driving alleles can benefit by distorting meiosis, individual selection generally favors Mendelian segregation (\citealt[e.g. ][]{Eshel1985}, but see \citealt{Ubeda2005,Haig2010}), creating a conflict between drivers and unlinked loci in the genome (i.e. `the parliament of genes,' \citealp{Leigh1971}).

Gametic drivers exploit the system of Mendelian segregation by providing a transmission advantage to their chromosome. 
Higher recombination rates make the ultimate chromosomal context of an allele uncertain, which can prevent the evolution of drive systems \citep{Thomson1974,Charlesworth1978,Haig1991}. 
It is therefore thought that modification of the recombination rate can evolve as a mechanism to alter the efficiency of gametic drivers. 
Conceptually, this model holds for both male and female drive systems.
However, gametic drivers are often sex-limited and display sex differences in the mechanisms by which they operate \citep[see][]{Ubeda2005}. 
Currently, the implications of sex differences in meiotic drive for the evolution of sex-specific recombination rates are unclear \citep[but see][ for an hypothesis]{Haig2010}.

Male gametic drivers (e.g. Segregation Distorter in \emph{Drosophila} and the t-haplotype in mice) usually operate after meiosis and are characterized by a two-locus damage-insensitive system \citep{Wu1991}. 
When these loci are tightly linked, a damage-insensitive haplotype can increase in frequency, even if this haplotype decreases individual fitness (e.g. \citealp{Prout1973, Charlesworth1978}). 
If the drive system imposes a cost, unlinked recombination enhancers can be favored for their ability to disrupt the drive system \citep{Thomson1974,Haig1991}. 
However, modifiers of the recombination rate linked to and in phase with driving alleles can increase in frequency if they decrease the recombination rate between components of a drive system  \citep{Thomson1974,Charlesworth1978}. 
This later idea is supported by the observation that gametic drivers in males are often located in inversions \citep{Wu1991}, which act to suppress recombination locally.

\begin{figure}
\begin{center}
\includegraphics{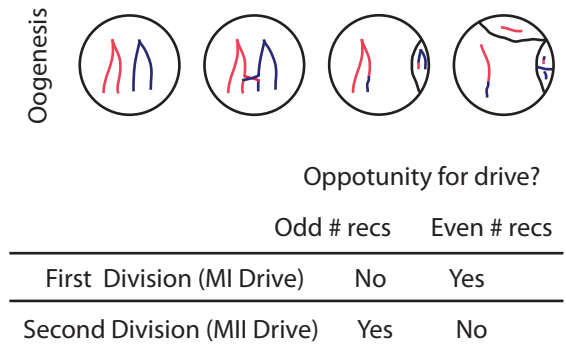}
\end{center}
\caption{{\bf Recombination during oogenesis and the opportunity for meiotic drive.}
Only one of the four products of female meiosis is included in the egg.
Recombination is a critical determinant for the opportunity for drive because it partitions variation within and among the products of the first meiotic division (dyads). 
With recombination between the marker and centromere, there is variation within but not among dyads, presenting an opportunity for drive during MII but not MI.
When recombination occurs after the marker, there is variation among but not within dyads, presenting an opportunity for drive during MI but not MII.}
\label{DriveCartoon}
\end{figure}

Female meiosis creates a single haploid product, providing an opportunity for alleles to compete for representation in the egg (true meiotic drive \citealp{Sandler1957,Zwick1999, Pardo-ManuelDeVillena2001a, Pardo-ManuelDeVillena2001b}). 
Recombination plays a fundamental role in female drive systems, as it determines at which stage of meiosis alleles can compete with each other. 
Since non-sister centromeres segregate at meiosis one (hereafter MI), an allele that biases the outcome of MI in favor of its centromere becomes over-represented in oocytes, so long as there is no (or an even number of) recombination (events) between driver and centromere (see Figure \ref{DriveCartoon}A.1 and 2A.2). 
The best-characterized cases of MI female drive are a subset of Robertsonian translocations in mammals \citep{Pardo-ManuelDeVillena2001c}, and a centromeric allele in \emph{Mimulus guttatus} \citep {Fishman2005,Fishman2008}.
Malik and Henikoff \citep[e.g. ][]{Malik2002,Malik2009,Malik2009a} have argued for a broad role, throughout eukaryotic evolution, of female centromeric drive at MI in driving the rapid evolution of centromeric sequence and the proteins that bind them.

Centromeres cannot drive during meiosis II (hereafter MII), as they are paired with their sister chromosomes. 
However, with a single recombination (or odd number of) event(s) between a focal locus and the centromere, the products of the first meiotic dividsion (dyads) will be variable at that locus, providing an opportunity for MII drive (see Figure \ref{DriveCartoon}). 

Examples of MII drivers include the Om locus in mice \citep{Wu2005} and Ab-10 in maize \citep{Rhoades1966}. 
The structure of the Ab-10 haplotype (an inversion spanning multiple loci) highlights the importance of recombination in the evolution of female drivers. 
A combination of the alleles in the Ab-10 system allows the chromosome to drive during MII \citep{Rhoades1966,Dawe1996}, while a distinct allele at another locus in this complex alters the recombination rate between itself and its centromere  \citep{Hiatt2003}, maximizing its ability to drive \citep{Buckler1999}.

In this manuscript we show that sex differences in meiotic drive can favor the evolution of heterochiasmy. 
With female meiotic drive, female recombination modifiers (unlinked to drivers) are favored when they decrease the efficacy of female drive. 
With segregating MI drivers, this corresponds to a female recombination enhancer, while female recombination suppressors decrease the efficacy of MII drivers. 
By contrast, when recombination modifiers and drivers are in one tightly linked haplotype, female recombination modifiers that increase the efficacy of drive are favored (Table 1).

\subsection{Relation to previous models}
Numerous authors have explored cases in which recombination modifiers are favored for their ability to break up systems of transmission ratio distortion \citep[e.g. ][]{Thomson1974,Charlesworth1978}.
\citet{Haig1991} pointed out that in addition to breaking apart drive systems, recombination can decrease the efficiency of drive by making the identity of an allele's partner in a dyad uncertain.

The potential role of meiotic drive in the evolution of heterochiasmy has received less attention. 
\citet{Lenormand2003} and \citet{Lenormand2005} briefly discussed male gametic drive systems, as a special case of the haploid selection model that may favor the evolution of heterochiasmy. 
As haploid chromosomes drive, the female meiotic drive model could also be considered as a special instance of the haploid selection model. 
However, we present our model as a distinct hypothesis because of its focus on recombination as a mechanism to modify the efficacy of meiotic drivers.

Recently, \citet{Haig2010} found that unlinked modifiers of the female recombination rate increase  in frequency when they \emph{enhance the efficacy of drivers} (the opposite of our finding, below). In both our model and Haig's, drivers decrease individual fitness. However, under Haig's model, the cost is reflected in the genetic identity of the products of meiosis, in which the fertility of a female heterozygous for a driver is determined by the genetic composition of her dyads. Haig's model equilibrates when the cost of drive on fertility is balanced by the degree of distortion. At this equilibrium, a modification of the recombination rate which increases female fertility also results in an increase in frequency of the driver, leading to the counterintuitive result that recombination modifiers unlinked to driver benefit by increasing the efficacy of drive. 
In other words, in Haig's model, females produce a higher proportion of viable gametes by creating chromosomal configurations that lessen the ill effects of drive. 
Somewhat paradoxically, by doing so, unlinked modifiers act to increase the frequency of drivers, despite the long term population-level fitness consequences of the spread of drivers.

In the model described below, we come to the opposite conclusion.
The discrepancy between our models arises from different conceptions of how drive influences fitness. 
In our model an individual's viability depends on it genotype at the drive locus, but the viability of all dyads are equivalent.
Therefore, the recombination modifier increases in frequency by decreasing the strength of drive.
By contrast, in Haig's model, recombination directly influences female fertility by influencing the viability of eggs.
Which model is more relevant to the evolution of heterochiasmy will depend on the mechanistic underpinning of the the cost of female drive.

\begin{table}
\label{predictions}
\begin {tabular} { | p{1.cm} | p{1.5cm} |  p{2.75cm} | p{2.25cm}  |p{7cm}  |}
\hline

Case & When & Linkage & Prediction & Reason \\  \hline
1 & MI & Unlinked, or repulsion phase & $r_\Venus > r_\Mars$ &  Female recombination enhancers discourage drive and hitchhike with high fitness, non-driving haplotypes. \\  \hline
2 & MI & Coupling & $r_\Venus < r_\Mars$ &  Female recombination suppressors hitchhike with driving haplotypes. \\  \hline
3 & MII & Unlinked, or repulsion phase & $r_\Venus < r_\Mars$ &  Female recombination suppressors discourage drive and hitchhike with high fitness, non-driving haplotypes. \\  \hline
4 & MII & Coupling phase & $r_\Venus > r_\Mars$ &  Female recombination enhancers hitchhike with driving haplotypes. \\  \hline

\end {tabular}
\caption{Relative male ($r_\Mars$) to female ($r_\Venus$) recombination rates as predicted by the meiotic drive theory.} 
\label{}
\end{table}

\section{The models}

Since neutral and beneficial drivers will quickly sweep through a population, and will not present a genetic conflict, we focus on the evolution of drivers that decrease organismal fitness.
We explore two models of drive. 
In the first model, the drive system consists of a single-locus that drives at either MI or MII. 
In our second model, drive is achieved by a two-locus system in which the ability of an MI centromeric driver to distort meiosis depends on the genotype at a partially linked drive enhancer locus.

For the single-locus drive model, we only care whether sister alleles are or are not separated during the first meiotic division (i.e. if there there is an odd ($1,~3,~5,\cdots$) or even ($0,~2,~4,\cdots$) number of recombination event occur between the centromere and the drive locus) in drive heterozygotes.
In our model, sisters are separated at the first meiotic division with probability $r$, or remain united with probability, $1-r$.

With an even number of recombination events between drive and centromeric loci, there will be variation within, but not among dyads. 
This genetic difference among dyads presents an opportunity for drive during MI, but prevents drive during MII. 
By contrast, an odd number of recombination events between drive and centromeric loci produces genetically homogenous dyads, precluding the possibility of drive during MI, but facilitating MII drive.
Given an opportunity for drive, drivers are represented in $\alpha_{\text{M}i} > 1/2$ of the gametes produced by drive heterozygotes. 
Without an opportunity for drive, meiosis is fair.

We then introduce another layer of biological complexity - the dependence of MI drive on the two-locus genotype at centromeric and drive enhancer loci.
In this two-locus system, the genotype at the drive enhancer locus influences the ability of the driving centromeric allele to distort segregation at MI. 
Without a drive enhancer, meiosis is fair.
By  contrast, the driving centromere is present in $\alpha_1$ of gametes from double heterozygotes, and $\alpha_2$ of drive enhancer homozygotes heterozygous at the centromeric drive locus.
As MII drive is unlikely to depend on the genetic identity of the centromere, we do not explore the case of two-locus drive during meiosis II.

Under this two-locus model, the outcome of multiple recombination events is somewhat complicated. 
Therefore, we make the restrictive assumption that only zero or one recombination events occur on a tetrad between the centromere and the drive locus (i.e. complete crossover interference). 
Thus, in our two-locus model $r$ is the probability of a single recombination event between the centromere locus and the drive locus, $1-r$ is the probability of no recombination, and we assume no double crossovers.

For each model, we contrast the case of sex-limited drive to results from a model in which both sexes drive. 
In these models, we introduce an allele that modifies the recombination rate between driver and centromere, without otherwise influencing organismal fitness. 
For models of single-locus drive, we compare the evolution of recombination modifiers in tight linkage with drivers to the evolution of recombination modifiers unlinked to drivers.

\section{Single-locus drive}

Consider a biallelic locus at which the driving allele, D, occurs in frequency, $f_{D}$, and the alternative allele occurs in frequency $f_{d} = 1 - f_{D}$. 
As we assume that drive has a pleiotropic cost on individual fitness, heterozygotes and drive homozygotes have fitnesses $w_{Dd} \leq 1$ and $w_{DD} <1 $, respectively. 
Although we assume that this cost in suffered equally by both sexes, allowing for sex-specific costs does not change the qualitative picture. 

The effective strength of drive acting in the $i^\text{th}$ meiotic division, $X_{\text{Mi}}$, is a function of both the recombination rate, $r$, and the strength of distortion, $\alpha_{\text{Mi}}$. 
With drive in both sexes, $X_{\text{MI}} = \alpha_{\text{MI}}(1-r)+r/2$, and $X_{\text{MII}} = r \alpha_{\text{MII}} + (1 - r)/2 $, for MI and MII drivers, respectively. 
With female-limited drive, the strength of MI and MII drive in oogenesis is 
$X_{\text{MI}\Venus} = \alpha_{\text{MI}\Venus}(1-r_\Venus)+r_\Venus/2$, and $X_{\text{MII}\Venus} = r \alpha_{\text{MII}\Venus} + (1 - r_\Venus)/2 $, respectively. 
Note that the effective strength of MI drive decreases with the recombination rate, while the effective strength of MII drive increases with the recombination rate. 
Note further that the effective strength of female drive is independent of the male recombination rate.

When drive has equivalent fitness effects in males and females, the mean population fitness, $W$, equals
\begin{equation}   W = f_{dd} + f_{Dd} w_{Dd} + f_{DD} w_{DD}\label{W}\end{equation}
where $f_\bullet$ denotes genotypic frequencies in newborns.
If drive operates in both sexes, the frequency of a driver after selection and drive is equivalent in sperm and eggs, and equals
\begin{equation} f_{D\mbox{M\emph{i}}}^{\prime} = \frac{f_{DD} w_{DD} + X_{\text{Mi}} f_{Dd} w_{Dd}}{W} \label{DeltaDriver} \tag{2.a} \end{equation}
When drive is female-limited, the recursive equation describing the frequency of drivers in sperm after selection ($f_{D\Mars}^{\prime}$) is
\begin{equation} f_{D\Mars}^{\prime} = \frac{w_{DD} f_{DD} + w_{Dd} f_{Dd} / 2}{W} \label{DeltaDriverSpermFemaleDrive} \tag{2.b} \end{equation}
and the recursive equation describing the frequency of drivers in eggs ($f_{D\Venus}^{\prime}$) is
\begin{equation}
f_{D\Venus\mbox{M\emph{i}}}^{\prime} = \frac{w_{DD} f_{DD} + w_{Dd} f_{Dd} X_{\text{Mi}\Venus} }{W}\label{DeltaDriverEggMIFemaleDrive} \tag{2.c}     \end{equation}
where genotypic frequencies equal
$ f_{DD}=f_{D\Venus}f_{D\Mars},
f_{Dd}=f_{D\Venus}f_{d\Mars}+f_{d\Venus}f_{D\Mars}$, and
$f_{dd}=f_{d\Venus}f_{d\Mars}$, and the subscripts, $\Venus$ and $\Mars$, represent allele frequencies in female and male gametes, respectively (i.e. egg and sperm). 
With drive in both sexes, allele frequencies are identical in sperm and eggs, and therefore genotypes are in Hardy-Weinberg Equilibrium; however, with female-limited drive, Hardy-Weinberg assumptions are inappropriate. 

In our appendix, we derive equilibrium frequencies of drive alleles in the absence of recombination modifiers by solving for $f_D$ when $\Delta f_D = 0$.
Since a protected polymorphism at the drive locus facilitates the evolution of recombination modifiers, the existence of these equilibria is important for the models below.

\subsection{Single-locus drive - An unlinked recombination modifier.}
We now investigate the coevolution of alleles at the drive locus and alleles at a locus which influences the recombination rate.
At the recombination modifier locus, alleles M and m occur in frequencies $f_M$ and $f_m = 1 - f_M$, respectively. 
The M allele additively alters the rate of recombination between driver and the centromere, has no direct effect on individual fitness, and is  unlinked to the drive locus. 
The recombination rate between drive and centromeric loci is $r$, $r+\delta r$, and $r+2 \delta r$ in mm, Mm, and MM individuals, respectively. 
In the case of drive in both sexes, we allow M to have equivalent effects in males and females.
With single-locus, female-limited drive we  only consider the influence of M on the female recombination rate, since our results are unaffected by $r_\Mars$.
We note that since M and D are unlinked, M does not influence the rate of recombination between itself and the drive locus.

The population consists of four haplotypes (md, Md, mD, and MD). Equations describing haplotype frequencies after selection, recombination, and drive, as well as many other derivations are presented in our appendix. 

To find the frequency of the recombination modifier, M, after one generation, we sum the frequencies of both haplotypes containing the M allele. 
With drive in both sexes, the change in frequency of M after selection, recombination, and drive equals
\begin{equation}
\Delta f_M = -\frac{\text{LD}}{W} (f_d (1- w_{Dd}) + f_D (w_{Dd} - w_{DD}))
\label{DMBS} \tag{3.a}
\end{equation}
where LD is the linkage disequilibrium between driving and recombination modifier alleles (i.e. LD $= f_{MD}-f_M f_D$), 
and is measured in the gametes which united at random to form this generation.

Equation \eqref{DMBS} shows that a genetic association between drive and modifier loci ($LD\neq 0$) is necessary for change in modifier frequency. 
More specifically, if we assume that the driver is costly and this cost is less than fully dominant (i.e. $w_{DD}\leq w_{Dd}\leq 1$, and $w_{DD} < 1$),  an unlinked recombination modifier increases in frequency when it is underrepresented in drive haplotypes (i.e. $LD <  0$), as it avoids the fitness cost accrued by driving alleles.

Analysis of the case of female-limited drive yields a similar conclusion. With sex-limited drive, the change in modifier frequencies in sperm and eggs equals
\begin{equation}        \Delta f_{M\Mars \mbox{MI}} =        \Delta f_{M\Mars \mbox{MII}} = \frac{f_{M\Mars}-f_{M\Venus}+z/W}{2}        \label{DSperm} \tag{3.b} \end{equation}
\begin{equation}        \Delta f_{M\Venus \mbox{MI}} =        \Delta f_{M\Venus \mbox{MII}} = \frac{f_{M\Venus}-f_{M\Mars}+z/W}{2}        \label{DEggs} \tag{3.c} \end{equation}
respectively, where 
\begin{equation} z =
(w_{Dd}-1) (\text{LD}_\Mars f_{d\Venus}+\text{LD}_\Venus f_{d\Mars})
+  ( w_{DD}-w_{Dd})(\text{LD}_\Mars f_{D\Venus}+\text{LD}_\Venus f_{D\Mars}) \label{z} \tag{4} \end{equation}

The first term in equations \eqref{DSperm} and \eqref{DEggs}, (i.e. $f_{M\Mars}-f_{M\Venus}$, and $f_{M\Venus}-f_{M\Mars}$, respectively) captures the role of syngamy in homogenizing allele frequencies across the sexes. 
The second term, $z/W$, is the change in modifier frequency due to linked selection on drive alleles. 
Since we assume that drive entails a less than dominant fitness cost (i.e. $w_{DD}\leq w_{Dd}\leq 1$, and $w_{DD} < 1$), $z$ will be positive when LD is negative. 
Therefore, recombination modifiers will increase in frequency when they are underrepresented on the driving haplotypes, like the case above. 
Additionally, since $z/W$ has the same role in changing modifier frequency ($\Delta f_M$) in male and female gametes, this model does not generate a sex difference in modifier frequency.
 
Since negative LD between recombination modifier and drive alleles is necessary for the adaptive evolution of the recombination rate, the salient question is - does our model generate negative LD? 
To address this question, we begin with a population in linkage equilibrium (LD = 0), and investigate the level of LD in gametes after selection, recombination, and drive. 

Starting from $LD=0$, the LD generated in a single generation with MI drive in both sexes equals
\begin{equation}
LD^{\prime}_{\mbox{MI}}=\delta r f_D f_d f_M f_m w_{Dd}(1-2\alpha_{\text{MI}}) / W  \tag{5}  \label{LDA}
\end{equation}
The LD generated with MII drive in both sexes equals 
$LD^{\prime}_{\mbox{MII}} = -LD^{\prime}_{\mbox{MI}}$. 
The LD generated between a female-limited driver and a recombination modifier after a single generation of selection, recombination, and drive is equivalent to the case described above replacing $\delta r$ and $\alpha$ by $\delta r_{\Venus}$ and $\alpha_{\Venus}$.

Note that with MI drive, negative LD is created between driving and recombination enhancing alleles (Equation \eqref{LDA} is less than zero when  $\delta r$ and $\delta r_\Venus$ are positive).
By contrast, with MII drive, negative LD is created between driving and recombination suppressing alleles (when $\delta r$ and $\delta r_\Venus$ are negative).
Since negative LD between driver and modifier results in an increase in modifier frequency (e.g. Equations (3.a) and (4)), recombination enhancers increase in frequency by hitchhiking with high fitness non-driving haplotypes.
Similarly, recombination suppressors are favored with MII drive.

We complement our single generation view of the creation of LD by making use of the Quasi Linkage Equilibrium (QLE) method
(e.g. \citealp{Kimura1965,Nagylaki1976,Barton1991,Kirkpatrick2002}). QLE relies on the fact that when linkage is loose, and selection and drive are weak, LD approaches an equilibrium value on a much faster time-scale than the slow change in allele frequencies. 
To derive this QLE LD, we solve for the equilibrium level of $\text{LD}^*$ (i.e. the LD for which $\Delta$ LD = 0), while holding allele frequencies constant.
To make this solution analytically tractable, we further assume that the strength of drive ($\alpha -1/2$ and $\alpha_\Venus -1/2$), selection ($1-w_{Dd}$ and $1-w_{DD}$), and that the degree of recombination modification ($\delta r$ and $\delta r_\Venus$), are all on order $\xi$, and we ignore terms of higher order than $\xi^2$. Doing this we find that 
\begin{equation}
\text{LD}^* \approx 2 \delta r f_d f_D f_m f_M (1 - 2 \alpha_{\text{MI}}) \tag{6.a} \label{QLE1}
\end{equation}
Comparing equation \eqref{QLE1} to equation \eqref{LDA} shows that much the equilibrium LD is generated in a single generation. Recalling that by definition, $\alpha_{\text{MI}} > 1/2$, it is clear that a recombination enhancer will be in negative LD with the drive allele, and will therefore be selectively favored. When considering MII drive, the result is conceptually similar but of opposite sign - with MII drive recombination suppressors, rather than enhancers generate negative LD with drivers.
 
Employing QLE methods, we can also derive the equilibrium LD between recombination modifiers and female-limited drivers, with the same assumptions as above. In this case, the equilibrium LD in sperm is a function of the LD in eggs, $\text{LD}_\Mars^* = \text{LD}_\Venus^* / 3$. 
Combining this result and our finding that $f_{M\Venus}=f_{M\Mars}$ , and assuming weak selection, the equilibrium LD in eggs with MI drive is approximately
 \begin{equation}
LD_\Venus^* \approx  \frac{3 \delta r (1-2 \alpha)f_{\textit{Dd}}f_{\textit{Mm}}}  {4 - (1-2 \alpha) (1 - r_\Venus)(f_D - 2 f_{d\Mars} )} \tag{6.b}
\end{equation}
Again, the QLE results show that most of the LD in females generated in a single generation. 
Note that, although selection does not directly generate LD in males (above), the inheritance of haplotypes from females maintains a low equilibrium level of LD in males. 
  
In summary, when unlinked to a female meiotic driver, a female recombination modifier can spread by reducing the ability of the driver distort meiosis in its favor \citep{Thomson1974, Haig1991}. 
By decreasing the strength of drive, the recombination modifier becomes under-transmitted on the drive haplotype, as the driving allele drives less efficiently in the presence of the modifier. 
Therefore the modifier hitchhikes with the high fitness, non-driving allele; however, unlike traditional models of hitchhiking, the modifier essentially arranges a ride for itself by increasing the expected transmission of the nondriving allele.

Furthermore, when drive is sex-limited, only recombination in the driving sex can act to decrease the efficiency of the single-locus drivers described above. 
\emph{
Therefore, a modifier with equal and opposite effects of male and female recombination rates can spread, 
and so our results meet the criteria of \citet{Lenormand2003} for the evolution of heterochiasmy. }

\begin{figure}
\includegraphics[width=14cm]{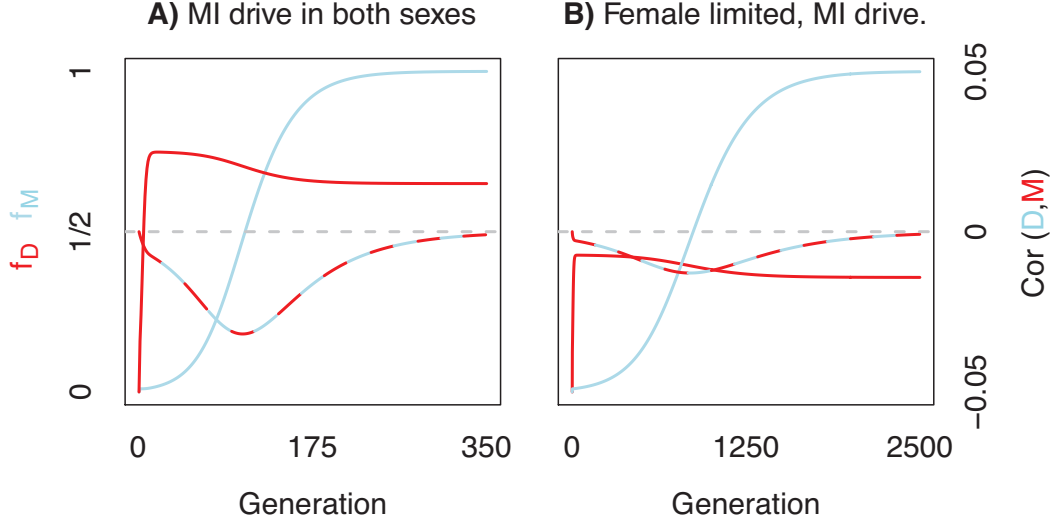}  
\caption{{\bf The coevolution of an MI driver and an unlinked recombination enhancer.}
The frequencies of MI drive alleles \textcolor{red}{($f_D$, red)}, and unlinked recombination modifiers {\color{SkyBlue}($f_M$, blue)} across generations. 
The correlation between alleles, $\frac{LD}{\sqrt{f_Df_df_Mf_m}}$, denoted by the  \textcolor{red}{red} {\color{SkyBlue} blue} line, and its value is given on the right axis.
Drive is complete and recessive lethal ($w_{Dd}=1$, $w_{dd}=0$). 
The initial recombination rate is 1/4, and each copy of M increases the probability of recombining by 0.05.
Initial frequencies of drive and recombination modifier alleles equal $f_{D_0}=0.10$ and $f_{M_0}=0.01$, respectively. 
{\bf \ref{Unlinked}A)} Drive in both sexes ($\alpha_{\text{MI}}=1$, $r=1/4$, $\delta r = 0.05$).
{\bf \ref{Unlinked}B)} Female-limited drive ($\alpha_{\text{MI}\Venus}=1$,  $\delta r_\Venus = 0.05$).
}
\label{Unlinked}
\end{figure}

Predictions concerning the evolution of recombination modifiers unlinked to MI and MII drivers are summarized by cases 1 and 2 in Table 1, respectively. 
Enhancers of the female recombination rate are favored when unlinked to female-limited MI drivers (case 1), while female recombination suppressors are favored when unlinked to female-limited MII drivers (case 3).
We display the dynamics of the case of the coevolution of a MI driver and an unlinked recombination modifier in Figure \ref{Unlinked}.

\subsection{Single-locus drive - A linked recombination modifier.}
 
We now explore the fate of a recombination modifier in tight linkage with the drive locus (i.e. there is no recombination between drive and recombination modifier loci). 
Intuitively, a modifier linked to a driver faces two opposing pressures - the deleterious effect of drive on individual fitness, and the selfish effect of drive on allelic transmission. 

These two components are captured by the equation describing the change in modifier frequency when drive operates in both sexes: 
\begin{equation} \Delta f_{M\mbox{MI}} = \frac{-LD(f_d (1- w_{Dd}) + f_D (w_{Dd}-w_{DD}))}{W} +\frac{LD\:w_{Dd} (1 - (r+\delta r))}{W}\label{DMBSlinkedMI} \tag{7.a} \end{equation}
\begin{equation} \Delta f_{M\mbox{MII}} = \frac{-LD(f_d (1- w_{Dd}) + f_D (w_{Dd}-w_{DD}))}{W} +\frac{LD\:w_{Dd} (r +\delta r)}{W}\label{DMBSlinkedMII}      \tag{7.b} \end{equation}
The change in frequency of a recombination modifier by individual selection is represented by the first term of equations \eqref{DMBSlinkedMI} and \eqref{DMBSlinkedMII} and is equivalent to the unlinked case (equation \eqref{DMBS}). 
As in equation \eqref{DMBS}, this term is positive when LD is negative.
The later term in equations \eqref{DMBSlinkedMI} and \eqref{DMBSlinkedMII} represents the change in frequency of the modifier due to drive, which is positive when LD is positive. 
Thus, although individual selection favors linked recombination modifiers in negative LD with drivers, transmission distortion favors modifiers in positive LD with drivers.

With female-limited drive, the intuition is similar; however, the recursive equations are more complex. The change in frequency of a recombination modifier in sperm, $ \Delta f_{M\Mars}$, is determined entirely by individual selection, and equals $ \frac{1}{2}(f_{M\Mars}-f_{M\Venus}+z/W)$, where $z$ retains its value from equation \eqref{z}.  The change in the frequency of a recombination modifier in eggs ($\Delta f_{M\Venus}$) equals
 
\begin{equation} \Delta f_{M\Venus \mbox{MI}} = \frac{1}{2}(f_{M\Venus}-f_{M\Mars} + \frac{z}{W} + \frac{u\:w_{Dd}}{W} (1 - (r_\Venus + \delta r_\Venus))) \label{DMBSlinkedFemaleMI} \tag{7.c} \end{equation}
\begin{equation} \Delta f_{M\Venus \mbox{MII}} = \frac{1}{2}(f_{M\Venus}-f_{M\Mars} + \frac{z}{W}  + \frac{u\:w_{Dd}}{W}  (r_\Venus +\delta r_\Venus)) \label{DMBSlinkedFemaleMII}    \tag{7.d} \end{equation}
where
\begin{equation} u=LD_\Venus+ LD_\Mars + (f_{D\Venus} - f_{D\Mars}) (f_{M\Venus} - f_{M\Mars}) \tag{8} \end{equation}
 
Therefore, with either sex-limited drive or drive in both sexes, both negative and positive LD between recombination modifier and driver can contribute to an increase in modifier frequency. 
Like the unlinked case, selection can generate linkage disequilibrium in a population in which no LD existed previously (see appendix). 
However, since a novel mutation must arise on one haplotype, the LD formed by the mutational history of tightly linked loci  is more important than is the LD generated by selection. 

\begin{figure}
\includegraphics[width=14cm]{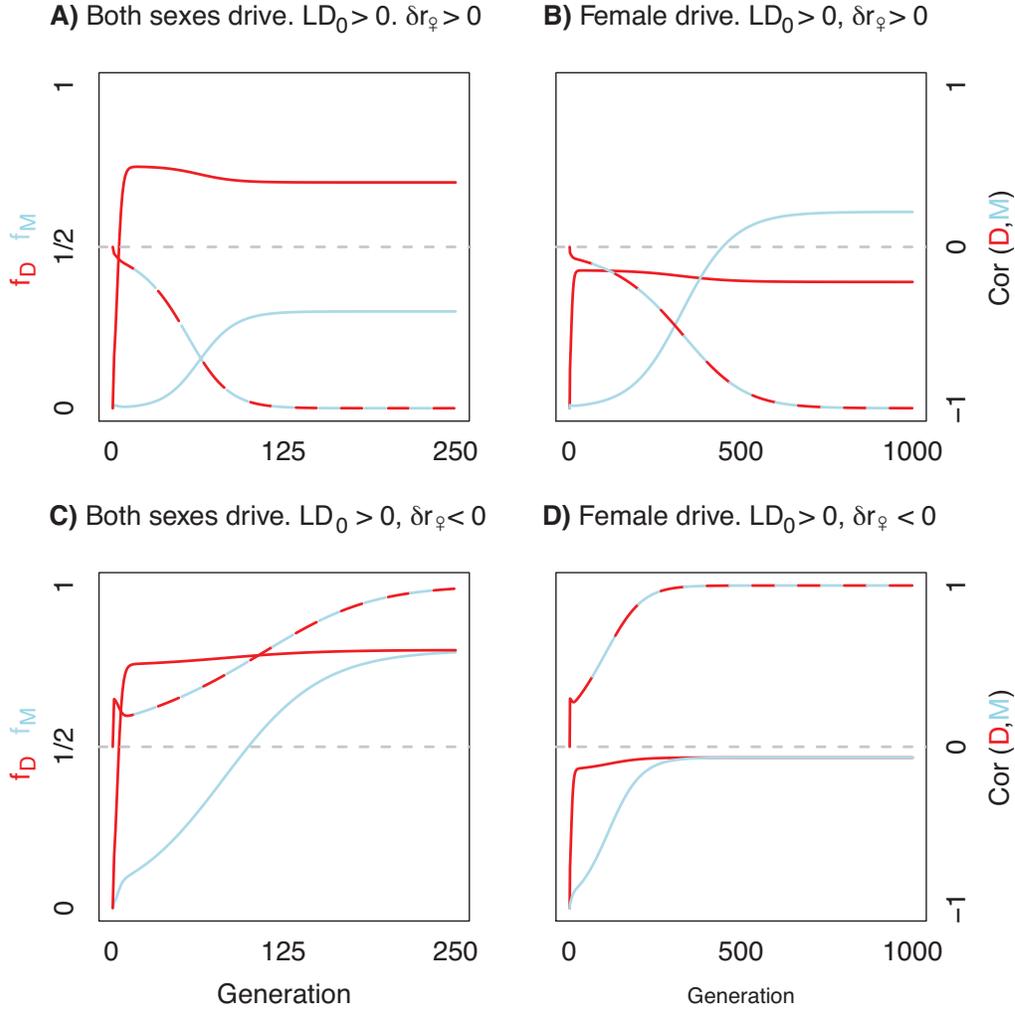}  
\caption{
{\bf  The evolution of drivers and recombination modifiers in tight linkage} 
The frequencies of MI drivers \textcolor{red}{($f_D$, red)}, and linked recombination modifiers {\color{SkyBlue}($f_M$, blue)} across generations. 
The correlation between alleles is denoted by the  \textcolor{red}{red} {\color{SkyBlue}blue} line, and its value is given on the right axis. 
Initial frequencies of driver and recombination modifier alleles are $f_{D_0}=0.10$ and $f_{M_0}=0.01$, respectively. 
Drive is complete and recessive lethal. Initial recombination rate equals 1/4.
{\bf \ref{Linked}A}) Drivers and recombination enhancement in both sexes ($ \delta r = 0.05$), 
M arises on a d chromosome. 
{\bf \ref{Linked}B}) Female-limited driver and recombination enhancement. ($ \delta r_\Venus = 0.05$), 
M arises on a d chromosome. 
 {\bf \ref{Linked}C}) Drivers and recombination suppression in both sexes. ($ \delta r = - 0.05$), 
 M arises on a D chromosome. 
{\bf \ref{Linked}D}) Female-limited driver and recombination suppression. ($\delta r_\Venus = -0.05$), 
M arises on a D chromosome. 
 }
\label{Linked}
\end{figure}

We display the population genetic dynamics of recombination modifiers linked to MI drivers in Figure \ref{Linked}.
Recombination enhancers increase in frequency when they arise on the non-driving background (Both sexes, Figure \ref{Linked}A. Female-limited, Figure \ref{Linked}B. See also  Table 1, case 1).
By contrast, recombination suppressors spread when they arise on the driving background (Both sexes, Figure \ref{Linked}C.  Female-limited, Figure \ref{Linked}D. See also Table 1, case 2).
The opposite result holds in the case of MII drive (Table 1, cases 3 and 4)

To provide a stronger intuition of the evolution of a recombination modifier linked to a drive locus, we investigate the special case of a recessive lethal driver which distorts meiosis in both males and females ($w_{Dd}=1$, $w_{dd}=0$, and $\alpha=1$). In this case, Equations \eqref{DMBSlinkedMI} and \eqref{DMBSlinkedMII}  become 
\begin{equation} \Delta f_{M\mbox{MI}}  = LD (1-f_D  - (r + \delta r))/W \tag{9.a} \label{reclethalmi} \end{equation} 
\begin{equation} \Delta f_{M\mbox{MII}} = LD (r+\delta r -f_D )/W \tag{9.b} \label{reclethalmii} \end{equation} 
respectively. 
Under the assumption of recessive lethality and complete drive, the equilibrium frequencies of MI and MII drivers is straightforward and equals $f_{D\text{MI}}^* =  1 - r $ and $f_{D\text{MII}}^* =  r $, respectively (see appendix). 
Plugging these values into equations \eqref{reclethalmi} and \eqref{reclethalmii} we find that the change in frequency of recombination modifiers in tight linkage with MI and MII drivers equals 
\begin{equation} \Delta f_{M\mbox{MI}} = - LD \delta r/W   \tag{9.c} \label{reclethalmib}  \end{equation}
\begin{equation} \Delta f_{M\mbox{MII}} =  LD \delta r/W  \tag{9.d} \label{reclethalmiib}  \end{equation}
respectively.  
Equations \eqref{reclethalmib} and \eqref{reclethalmiib} provide a straightforward characterization of the invasion of a rare recombination modifier.
With MI drive and tight linkage (equation \eqref{reclethalmib}) a recombination enhancer will increase in frequency when it arises on the non-driving haplotype $(LD < 0, \delta r > 0$, gives $\Delta f_{M\mbox{MI}} > 0$), 
while a recombination suppressor will increase in frequency when it arises on the driving haplotype $(LD > 0, \delta r < 0$, gives $\Delta f_{M\mbox{MI}} > 0$)
The opposite result holds for a recombination modifier tightly linked to an MII driver (Equation \eqref{reclethalmiib}).
We note that equations \eqref{reclethalmib} and \eqref{reclethalmiib} only hold for the first generation of selection.

The analysis of female-limited drive is more complex, but ultimately yields a similar result. 
Here, we present our invasibility analysis in which we derive results under female-limited drive.
In this case, female-specific recombination enhancers are favored when in repulsion phase with MI drivers or coupling phase with MII drivers, while the opposite results hold for recombination suppressors  (see `Invasibility analysis,' pages 18-19, 21 in our appendix). 
For this invasibility analysis (see \citealp{Otto2007}), we write the recursion for haplotype frequencies in sperm and eggs after selection, recombination, and drive in matrix form.

When eigenvalues of the Jacobian of this matrix (evaluated at viability-drive equilibrium setting the frequency of the recombination modifiers to zero) are greater than one, a rare recombination modifier increases in frequency. For both the cases of MI and MII drive, there are two eigenvectors of interest. One eigenvector is greater than one when $\delta r_\Venus >0$, the other is greater than one when $\delta r_\Venus <0$ suggesting that both recombination enhancers and suppressors can increase in frequency). Because the LD generated in these cases is large (in fact, at equilibrium there are only two haplotypes), we do not employ the QLE approach, which works best under loose linkage \citep{Barton1991,Kimura1965,Kirkpatrick2002,Nagylaki1976}.

Unlike the case of unlinked recombination modifiers, modifiers linked to drive loci do not generally approach fixation. Assuming that the modifier is favored, it rapidly goes to fixation on the background onto which it mutated; however, this haplotype now moves to its new equilibrium frequency determined by the new recombination rate. So long as this new equilibrium is greater than zero and less than one, recombination modifiers will be stably polymorphic (see Figure \ref{Linked}). 

In sum, the evolution of recombination modifiers linked to drivers yields a rich and diverse set of predictions. When tightly linked to MI drivers, recombination enhancers in repulsion phase (Table 1, case 1), and recombination suppressors in coupling phase are favored (Table 1, case 2). 
When tightly linked to MII drivers, recombination suppressors in repulsion phase (Table 1, case 3), and recombination enhancers in coupling phase are favored (Table 1, case 4). 
\emph{Like the unlinked model (above), the fate of a female recombination modifier linked to a female-limited driver is independent of its influence on the male recombination rate.
Thus, a modifier with equal but opposite effect on male and female recombination rates (i.e. no net effect) can spread,  facilitating the evolution of heterochiasmy.}

\section{Two-locus drive systems}

We now turn our attention to the more complex case of two-locus, MI drive. 
In this model the strength of drive by a centromeric variant, C, depends on the genotype at the drive-enhancer locus, D, which is on the same chromosome as the centromeric driver. 
Specifically, in Cc heterozygotes, meiosis is fair in a genetic background of d homozygotes, but C is represented in $\alpha_1$ and $\alpha_2$ of gametes from Dd/Cc and DD/Cc individuals, respectively (where $\alpha_2 \geq \alpha_1 \geq 0.5 $ and $\alpha_2>0.5$).  
Although it is possible that the drive enhancer will incur an individual fitness cost, we focus on the case in which the drive enhancer is neutral, but the driving centromere is costly. 
Imposing a fitness cost to drive enhancer adds subtle quantitative differences to the results, and this model where both loci involve costs has been well explored in the description of the SD system \citep{Hartl1975,Charlesworth1978,Haig1991}. 
Since the genetic identity of a centromere seems unlikely to influence MII drive, we do not pursue a two-locus model of MII drive.

With two-locus, MI drive, a recombination enhancer can increase in frequency, and ultimately approach fixation, as in the case of single-locus MI drive \citep{Thomson1974,Haig1991}. 
With drive in both sexes, the change in frequency of a recombination modifier is 
\begin{equation}
\Delta f_M = \frac{-\text{LD}_{MC}}{W}(f_c(1-w_{Cc})+f_C(w_{Cc}-w_{CC}))\label{3lm}
\end{equation}
where $\text{LD}_{MC}=f_{MC}-f_M f_C$ is the linkage disequilibrium between centromere and recombination modifier. 

\begin{figure}
\begin{center}
\includegraphics[width=17cm]{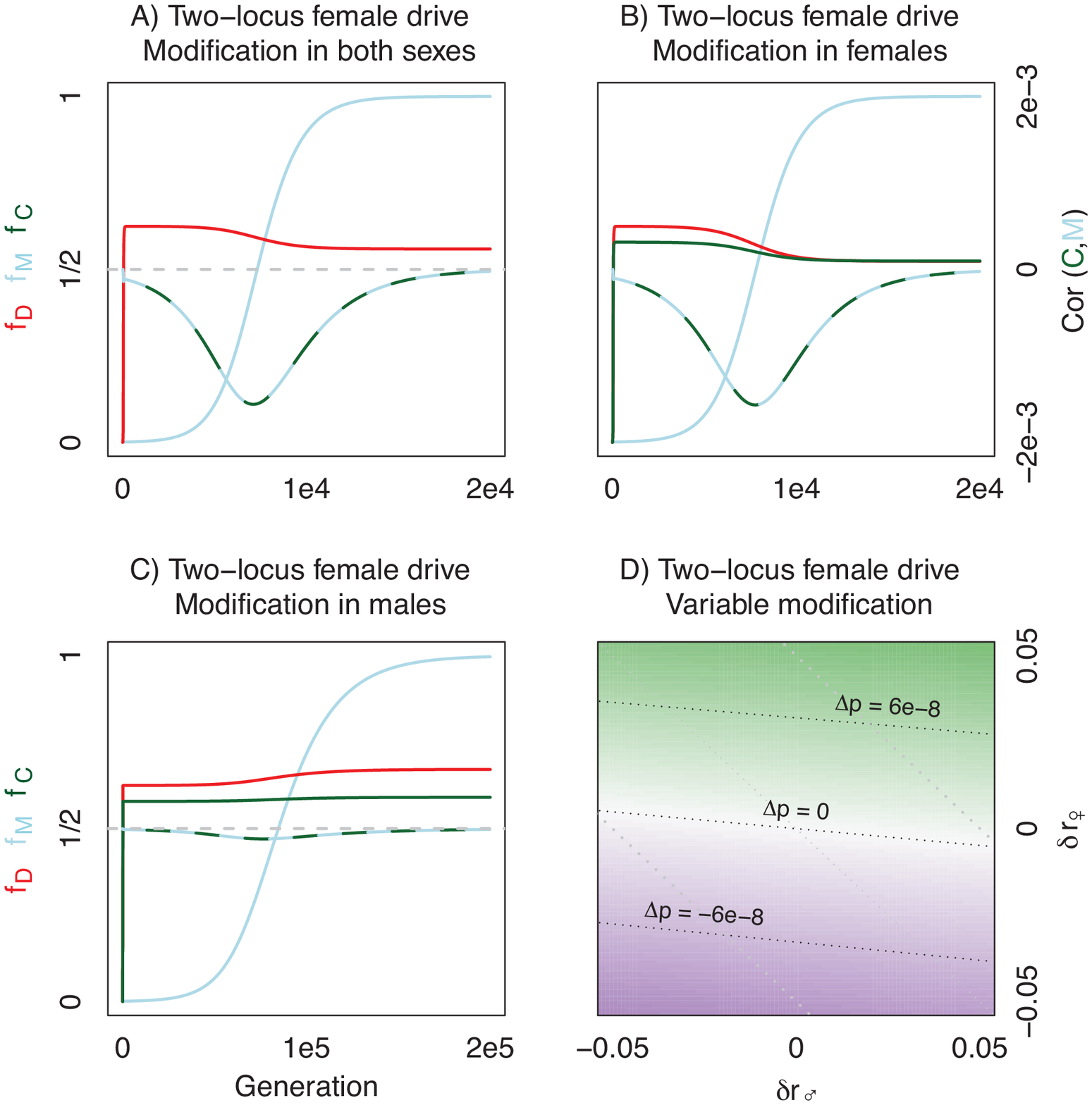}   
\end{center}
\label{3Locus}
\end{figure}

As in the case of recombination modifiers unlinked to single-locus MI drivers, the recombination enhancer spreads by becoming underrepresented in the low fitness, centromeric driving (C) genetic background. 
Recombination enhancers generate this LD by decreasing the expected transmission of the drive enhancer allele, which allows the recombination enhancer to escape from the driving haplotype (Figure \ref{3Locus}A).

When drive is female-limited, alleles that increase the recombination rate between drive enhancer and centromeric loci in either sex  are favored. 
However, female-limited recombination enhancers spread much more quickly and have more negative LD with driving centromeres than do male-limited recombination modifiers. 
For example, in Figure \ref{3Locus}B it only takes approximately 11,400 generations for a female-limited recombination modifier to rise from a frequency of 0.1\% to 95\%, but it takes more than an order of magnitude longer for a male-limited recombination enhancer to reach this frequency (Note the order of magnitude difference on the x-axis in  \ref{3Locus}B and  \ref{3Locus}C).

Because female recombination enhancers are more strongly favored than male recombination enhancers in this system, alleles that increase female recombination can increase in frequency, even if they drastically reduce male recombination rates (Figure \ref{3Locus}D). Therefore, our two-locus model also passes the `no-net-effect test' \citep{Lenormand2003} - facilitating the evolution of heterochiasmy.    

The intuition behind this numerical result is as follows. 
Drive in females generates a positive association between centromeric drivers (C) and linked drive enhancers (D). 
With this association, D increase in frequency during female (but not male) meiosis. 
Recombination in females decreases both the expected transmission of drive-enhancers, and the co-transmission  of D and C alleles. 
Since gametes from mothers with higher recombination rates have fewer D alleles than expected, drive in their daughters is less efficient. 
Ultimately, the granddaughters of females with higher recombination rates  suffer less from the deleterious fitness effects of drives systems \citep[e.g. ][]{Crow1991}.

As male recombination does not directly change the transmission of a female meiotic drive enhancer, the selection on a male recombination modifier is weak. 
Nonetheless, elevated male recombination rates do break down the association between D and C alleles, which ultimately allows male recombination enhancers to escape from centromeric drivers, providing a minor boost to male-limited recombination enhancers. However, this effect is meagre compared to the effect of female recombination modification (Figures \ref{3Locus}B-D).

\newpage
Figure 5: {\bf The co-evolution recombination modifiers and a two-locus drive system.} 
The frequencies of drive enhancer, centromeric driver, and recombination modifier alleles are displayed in \textcolor{red}{red}, \textcolor{green}{green}, and {\color{SkyBlue}blue}, respectively. 
The correlation between recombination modifier and centromeric driver alleles is denoted by the  \textcolor{green}{green} {\color{SkyBlue}blue} line, and its value is given on the right axis of Figure 5B (The scale is maintained in Figure 5C). 
Initial frequencies drive and recombination modifier alleles equal $f_{D_0}=0.10$ and $f_{M_0}=0.01$, respectively. 
The driving centromeric allele completely distort meiosis in DD and Dd genotypes (e.g. $\alpha_1=\alpha_2=1$), and is a recessive lethal. 
Neither drive enhancer nor recombination modifier directly influence individual fitness. 
The initial recombination rate and allele frequencies are:  $r=0.10$, and $f_{D(0)} = f_{C(0)} = f_{M(0)} =0.001$, respectively. 
{\bf A)} Recombination modification in both sexes ($\delta r_\Venus = 0.025$, $\delta r_\Mars=0.025$). 
{\bf B)} Female-limited recombination modification ($\delta r_\Venus = 0.025$, $\delta r_\Mars=0$). 
{\bf C)} Male-limited recombination modification ($\delta r_\Venus = 0$, $\delta r_\Mars=0.025$).
{\bf D)} The modifier has distinct influence on male and female recombination rates ($\delta r _\Mars$, and $\delta r _\Venus$, respectively). \textcolor{green}{Green} indicates an increase in modifier frequency, {\color{Purple} purple} 
indicates a decrease. Labels above diagonal lines describe the relative change in allele frequencies over 250 generations.

\newpage
\section{Discussion}

Meiosis and recombination are deeply conserved and highly orchestrated processes, in which slight errors can have severe fitness consequences. 
Nonetheless, many of the functional components of meiosis and recombination evolve rapidly \citep[e.g. ][]{Malik2002,Anderson2009,Myers2010}. 
One explanation for this rapid evolution is that meiosis and gametogenesis offer a number of opportunities for genomic conflict within an individual, generating a pattern of antagonistic coevolution between selfish gametic drivers and suppressors of meiotic drive \citep[see][ for a broad overview]{Burt2006}.

Since the progression of meiosis and gametogenesis is highly sex-specific \citep{Morelli2005}, we expect that forms of conflict and conflict mediation will also be sex-specific. 
The asymmetry in meiotic division during oogenesis presents an opportunity for competition between alternative alleles for representation in the egg \citep{Sandler1957,Zwick1999,Pardo-ManuelDeVillena2001b,Pardo-ManuelDeVillena2001a,Malik2009a}. 
Because recombination determines the ability of female drivers to distort meiosis at MI and MII, female recombination rate influences the ability of a driver to distort meiosis.
We have shown that female meiotic drive can favor changes in the female recombination rate -  female recombination modifiers are selected to enhance or suppress drive (see Table 1) with changes in male rate having little or no effect on the efficacy of female drive.

Our understanding of the frequency, severity, and operation of female meiotic drive systems is still in its infancy. 
Therefore, we have not based our population genetic analysis on explicit mechanistic details of female drive. 
However, our models can be related to biologically plausible mechanisms.
The model of a single-locus MI driver could correspond to an epigenetic modification of a centromere in \emph{cis}, or a structural stretch of DNA that influences the orientation of the centromere in such a way as to increase its probability of inclusion into the secondary oocyte. 
In our two-locus MI drive system, the centromeric locus could correspond to a centromeric satellite that increases its probability of inclusion in the primary oocyte through an interaction with the spindle, while alternative alleles at the drive enhancer locus could represent centromeric proteins which interact with the centromeric machinery to enhance or suppress the effect of drive \citep{Malik2002,Malik2009,Malik2009a}. 
Our MII model roughly corresponds to neocentromeric drive systems, such as the Ab-10 locus in maize \citep{Rhoades1966}, or telomeres that influence the orientation of meiotic chromosomes \citep[][see \citet{Anderson2008} for a discussion]{Novitski1951}.
We reiterate that our results depend solely on the ability of recombination to modify drive systems, rather than  on any specific drive mechanism.

We have shown that female meiotic drive systems create a selective environment that favors the evolutionary modification of the female recombination rate.
However, selection on female recombination rates does not necessarily lead to heterochiasmy. 
For example, in the extreme case where  if alleles that modify the recombination rate in females have equivalent effects on the male rate, selection on the female rate will not generate heterochiasmy. 
However, above we show that our model favors modification of the female recombination rate even if modifiers have opposing effects the on male rate - the standard for models of the evolution of heterochiasmy \citep{Lenormand2003}.

While we still know little about recombination modifiers, current evidence suggests that the genetic control of the global recombination rates differs by sex, and therefore selection on the female rate is likely to generate heterochiasmy.
Three distinct lines of evidence support this tentative conclusion.
First, the global control of meiosis and recombination is sexually dimorphic \citep{Morelli2005}.
Second, the few naturally occurring alleles known to modify the total genetic map length in humans do so in a highly sex-specific manner \citep{Kong2004,Kong2008,FledelAlon2011}.
Third, although there is additive genetic variation for the map length in both sexes, no heritable intersexual correlation in map length has been found \citep{FledelAlon2011}.

The predictions of our model are sensitive to biological details such as the linkage association between drivers and recombination modifiers, as well as the timing of drive (MI vs MII). 
This makes it hard to generate concrete predictions about whether female drive will select for higher or lower female recombination rates. Indeed, the fact that female rates are not always higher than male rates suggests that the direction of selection may not be constant.  
One concrete prediction is that since the centromere holds a special place in female meiotic drive (in both MI and MII systems) a constant influx of new female drivers will systematically select particularly for heterochiasmy in the regions surrounding centromeres.

The observation in many taxa of higher female recombination rates, especially near the centromere is consistent with two different biological scenarios elaborated in our models.
First, elevated female recombination rates, especially near centromeres, may represent  the action of unlinked suppressors to prevent the spread of MI female drivers.
Alternatively, this pattern could be explained by the spread of recombination enhancers linked to MII drivers, which increase in frequency because recombination enhancers facilitate MII female drive. Empirical progress in elucidating the genetic basis of local variation in sex-specific recombination rates and female meiotic drive across the genome will shed light on which (if any) of these two models explain this pattern.

In contrast to global modifiers of the recombination rate, we know a little more about local modifiers of recombination rate, though our picture is still far from complete. 
One broad class of \emph{local suppressors} are chromosomal inversions, which seem to be a common response to selection for reduced recombination \citep[see ][ for discussion]{Kirkpatrick2010}. 
Inversions are {\it a priori} expected to locally suppress recombination similarly in both sexes and this heterozygous effect will be removed when the inversion is eventually lost or fixed within the population. 
Therefore, given our current understanding of local recombination modification, we think it is unlikely that selection for linked suppressors of recombination will strongly contribute to heterochiasmy.

Although we focused on female-limited drive, there are well documented cases of male-limited transmission distortion involving multilocus gene complexes (Note that none of these are true meiotic drive, relying instead on sperm death)  \citep{Wu1991}.
A model of the coevolution of a two-locus male drive system and a recombination modifier will yield results similar to the case of two-locus MI female drive described above (with slight differences due to the fitness of recombinants in male systems). 
That is, with male-limited drive systems, we expect that recombination modifiers in coupling phase would benefit from reducing the male recombination rate, while unlinked modifiers would benefit from increasing the male recombination rate \citep[see also discussion by ][]{Lenormand2005}. 
Evidence for the former is bountiful, as most known male transmission ratio distorters are tied together by complex inversions \citep[e.g. ][]{Presgraves2009}. 
However, unlike female meiotic drivers, male distortion systems can arise at any chromosomal location irrespective of the distance to the centromere. 
Therefore, even if male distortion systems do select for heterochiasmy, a constant influx of new male systems will not systematically select for sex differences on a chromosomal-scale.

More generally, our models may explain other broad patterns associated with recombination. 
One outstanding pattern is the observation that variation in the number of chromosome arms is a better predictor of recombination rate variation among mammalian species than is variation in the number of chromosomes, or the physical size of the genome \citep{Pardo-ManuelDeVillena2001d}. 
This result is somewhat surprising because only one recombination event per chromosome is required for proper segregation, arms of metacentric chromosomes are often found to be lacking a crossover \citep{Fledel-Alon2009}, and the centromere seems to offer no barrier to interference in many systems \citep{Broman2000,Fledel-Alon2009,Demarest2011}.
The meiotic drive theory can explain these observations by proposing that modifiers of the recombination rate are selected to increase recombination events between centromeres and potential drivers, as both sides of a chromosome present an opportunity for drivers to exploit meiosis. 

Another broad pattern is the observation that heterochiasmy is reduced in selfing plants, which \citet{Lenormand2005} saw as support for their hypothesis of a role of haploid selection on male gametes in driving heterogamy. 
We note that this observation is potentially consistent with the female meiotic drive hypothesis - since most selfing plants are largely homozygous, there is little opportunity for drive. 
The alteration of generations in plants provides exciting opportunities for future research on the evolution of heterochiasmy.

The meiotic drive theory could also explain the observation of rapid changes in recombination rates \citep{Coop2007,Dumont2008,Dumont2011}, as the recombination map will constantly be evolving as recombination enhancers and suppressors respond to new drive systems across the genome. 
A greater knowledge of medium-scale patterns of turnover in male and female rates would help clarify the plausibility of meiotic drive in explaining this pattern. For example, the meiotic drive theory would be strongly supported if regions close to centromeres show particularly high rates of female recombination evolution, as is observed in the \emph{Drosophila} clade \citep{True1996}.

Further tests of these predictions require the ability to identify both meiotic drivers and modifiers of the recombination rate. 
Currently, our knowledge of the distribution of meiotic drivers and their fitness effects is very incomplete, and is likely biased towards the overrepresentation of strong drivers with extreme fitness effects. 
However there is mounting evidence supporting the existence of subtle transmission ration distorters  \citep[e.g. ][however, the mechanism of distortion in these cases is often unknown]{Reed2005,Zollner2004,Aparicio2010,Axelsson2010}. 
Similarly, although there is ample evidence that the recombination rate, as well as the strength and direction of heterochiasmy varies across species, few allelic variants that influence sex-specific recombination rates have been identified. 
Although we currently know very little about the frequency of female drivers, or the genetic control of sex-differences in the recombination rate across many taxa. Fortunately, as technological advances make the sequencing of offspring (or gametes) from many meioses more affordable, identification of alleles governing sex-specific transmission and recombination rates will become much easier.

More broadly, the ideas presented here are part of a larger body of the theoretical work that highlights the potential diverse role of genomic conflict in shaping the evolution of recombination, meiosis and gametogenesis (e.g. \citealp{Sandler1957,Haig1991,Haig2010,Hurst1996, Zwick1999,Malik2009,Anderson2009} \citealp{Burt2006}, Table 12.3). 
For example, conflicts between \emph{cis} and \emph{trans} determinants of hotspot localization due to bias gene conversion have been put forward to explain the rapid evolution of mammalian fine-scale recombination rates and their determinants (such as the hotspot binding protein Prdm9 \citep{Baudat2010,Myers2010,Parvanov2010}) \citep{Boulton1997,Coop2007b,Ubeda2011}.
While much of the classic work on the evolution of recombination and meiosis has focused on the benefits of creating adaptive gene combinations, purging deleterious recessive alleles from an adaptive haplotype, or bringing together two beneficial mutations onto one haplotype \citep[e.g. ][]{Eshel1970,Feldman1996,Otto2002,Barton2009}, it seems equally plausible that the short-term evolution of recombination rates may be in response to conflicts created during meiosis. 
\newpage

\section{Acknowledgements}
We thank Chuck Langley for discussions that inspired this paper. This paper was greatly improved by comments from David Haig, Chuck Langley, Pat Lorch, Molly Przeworski, Peter Ralph, Alisa Sedghifar, and Michael Turelli. We thank Thomas Lenormand and the other anonymous reviewer for very helpful feedback. This work was made possible by a NSF Bioinformatics postdoctoral fellowship to YB and support to GC from the Sloan Fellowship in Computational and Evolutionary Molecular Biology.

\end{document}